\documentclass[twocolumn,showpacs,prl]{revtex4}
\usepackage{graphicx}
\usepackage{dcolumn}
\usepackage{bm}

\begin{document}

\title{Quantum phase diagram of fermion mixtures
with population imbalance in one-dimensional optical lattices}
\author{B. Wang, Han-Dong Chen, and S. Das Sarma}
\affiliation {Condensed Matter Theory Center and Center for
Nanophysics and Advanced Materials, Department of Physics,
University of Maryland, College Park, Maryland 20742-4111}

\begin{abstract}
With a recently developed time evolving block decimation (TEBD)
algorithm, we numerically study the ground state quantum phase
diagram of fermi mixtures with attractive inter-species
interactions loaded in one-dimensional optical lattices. For our
study, we adopt a general asymmetric Hubbard model (AHM) with
species-dependent tunneling rates to incorporate the possibility
of mass imbalance in the mixtures. We find clear signatures for
the existence of a Fulde-Ferrell-Larkin-Ovchinnikov (FFLO) phase
in this model in the presence of population imbalance. Our
simulation also reveals that in the presence of mass imbalance,
the parameter space for FFLO states shrinks or even completely
vanishes depending on the strength of the attractive interaction
and the degree of mass imbalance.
\end{abstract}

\pacs{03.75.Ss; 67.85.-d; 71.10.Fd}

\maketitle

The pairing of fermions in multi-component fermi mixtures is of
fundamental interest not only in condensed matter, but also in
atomic, nuclear, and astro- physics. When the fermi surfaces of
the component species are mismatched due to unequal densities,
possibilities open up for exotic pairing mechanisms such as the
Fulde-Ferrell-Larkin-Ovchinnikov (FFLO) \cite{FF64,LO65} and the
breached pair (BP) states \cite{Sarma63,Liu03}. While these
interesting states of matter remain experimentally elusive,
controversies about their stability have been long-standing.
Recently, this subject has received renewed intense attention due
to atomic physics experimental studies on population-imbalanced
ultra-cold fermions \cite{Zwierlein06}. The advent of techniques
for precisely controlling and detecting ultra-cold atoms is making
it possible to systematically study the FFLO and/or the BP states
with exotic fermionic pairing, which has turned out to be
impossible to do in conventional solid state superconductors.

Inspired by the experimental achievements, over the past few years
a large body of theoretical studies have been reported for the
exotic pairing states in ultra-cold atomic fermion systems with
population imbalance \cite{3DFFLO,3DSarma,1DFFLO,1DFFLOb}. It has
been found that both FFLO \cite{3DFFLO} and BP states
\cite{3DSarma} could exist in 3D systems. Particularly, a
consensus has developed that the stability region of the 3D FFLO
states is very narrow, making it very difficult to observe
experimentally. In contrast, due to the Fermi surface nesting it
could be much easier to observe the FFLO-type states in 1D or
quasi-1D systems \cite{1DFFLO,1DFFLOb}. In this work we present
numerically calculated exact quantum phase diagrams for the
population imbalanced 1D fermi mixtures with unequal masses for
the component species.

Fermi mixture systems with mass imbalance could be characterized
by the asymmetric Hubbard model (AHM) with species dependent
tunneling rates. With equal spin populations, a phase diagram for
the 1D AHM has been obtained with the renormalization group
technique \cite{Cazalilla05} and possible spin segregation in this
model with repulsive interactions has been investigated with
bosonization and density matrix renormalization group (DMRG)
techniques \cite{Gu07}. Nonetheless, it is not surprising that
more attention has been given to the systems with both mass and
population imbalances, in which stabilities of FFLO and breached
pair (BP) states in 3D systems have been extensively
studied\cite{3Dmass}. Very recently, a quantum Monte Carlo (QMC)
study on finite-size fermi mixtures in 1D optical lattices with
imbalanced populations and masses has been
reported\cite{Batrouni08b}, in which evidence of 1D analog of
FFLO-type states has been found. The QMC simulation also reveals
that when the mass difference is large enough, instead of an
FFLO-type state, the ground state of the system will become an
inhomogeneous ``collapsed'' state.

In this paper, we report a time evolving block decimation (TEBD)
numerical study on fermi mixtures with unequal masses and
attractive on-site interaction in 1D optical lattices. Our study
is complementary to the QMC simulation in Ref. \cite{Batrouni08b},
in that we obtain the phase diagrams for such fermi mixture
systems. Consistent with the QMC results, we find that FFLO states
are the only possible class of polarized pairing states in such
systems. As the mass imbalance increases, the parameter space for
FFLO states shrinks and eventually vanishes completely. Since we
study the homogeneous system at the thermodynamic limit, our phase
diagram does not have any inhomogeneous collapsed state, in
contrast to Ref. \cite{Batrouni08b}.

\begin{figure} [tbp]
\includegraphics [height=6 cm, width=8.5 cm] {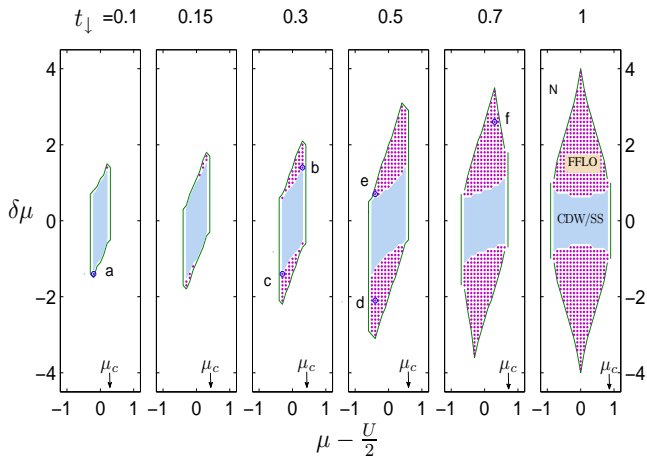}
\caption [Fig.1] {Effective magnetic field vs. chemical potential
ground state phase diagrams for $U=-4$. From left to right, the
six panels correspond to $t_{\downarrow}=0.1$, 0.15, 0.3, 0.5,
0.7, and 1, respectively. The ground states of the system in the
regions filled with solid dots are found to be of the FFLO type.
In the shaded regions around the center of each panel, no
population imbalance presents in the ground states. From
$|\mu-U/2|=0^{+}$ to $\mu_c$, the dominant/subdominant quasi-long
range orders changes from CDW/SS to SS/CDW. In the remaining
regions, at least one species of fermions satisfies $\langle
n_{\sigma i}\rangle$=0 or 1 and the system behaves as normal Fermi
gas.}
\end{figure}

For our simulation, we adopt the asymmetric Hubbard Hamiltonian to
model the Fermi mixtures in 1D optical lattices with unequal
masses:
\begin{eqnarray}
\nonumber H=&-&\sum_{\langle i,j\rangle, \sigma}
t_{\sigma}c^{\dagger}_{\sigma i}c_{\sigma j}+\sum_i Un_{\uparrow
i}n_{\downarrow i}\\&-&\sum_i \mu (n_{\uparrow i}+n_{\downarrow i}
)+\sum_i\delta\mu(n_{\downarrow i}-n_{\uparrow i}),
\end{eqnarray}
where $c_{\sigma i}$ ($\sigma=\uparrow$ or $\downarrow$) stands
for the annihilation operator of $\sigma$-fermion species at site
$i$, $n_{\sigma i}(=c^{\dagger}_{\sigma i}c_{\sigma i})$
represents the corresponding particle number operator,
$t_{\sigma}$ is the species-dependent tunneling rate (through out
this paper $t_{\uparrow}$ is set to be the unit of energy), $U$
characterizes the on-site inter-species interaction strength, and
$\mu\pm\delta\mu$ gives the chemical potential of the
$\uparrow$/$\downarrow$-fermion. In analogy with real spin
systems, $\delta\mu$ will be referred as effective magnetic field
in the following, although our system is spinless with the
$\uparrow/\downarrow$ components referring to fermions with
different masses. The unequal mass effects are incorporated in the
species-dependent tunneling rates through
\begin{equation}
\frac{t_{\downarrow}}{t_{\uparrow}}=\frac{m_{\uparrow}}{m_{\downarrow}}.
\end{equation}
We note that the asymmetric Hubbard model defined by Eq. (1) could
also be realized in mixtures of same-species fermions prepared in
two different internal states by engineering
internal-state-dependent optical lattices \cite{Liu04}.

To investigate the ground state properties of this Hamiltonian, we
use an infinite lattice version of the TEBD algorithm \cite{TEBD},
which allows us to study the model in the thermodynamic limit. A
source of intrinsic numerical error for TEBD is due to the
Trotter-Suzuki expansion (TSE) used in the decomposition of time
evolution operator. (In our simulation we choose the 4th order
symmetric TSE.) Furthermore, the convergence of the physical
results with TEBD is mainly controlled by a cut-off parameter
$\chi$, which characterizes how well one preserves the bipartite
entanglement of the system when truncating the Hilbert space. In
this work, we choose $\chi=60$. The convergence has been checked
to be good enough (within $\sim\mathcal{O}(10^{-4})$ for real
space correlation functions) for our purpose, as compared with
$\chi=80$ and 100 results.

To identify relevant (quasi-)phases for our model, we calculate
the real space spin-spin ($S_r^{m}$), density-density ($D_r$), and
pairing ($P_r$) correlations and their Fourier transforms
$X_{k}=1/\sqrt{M}\sum_{r=0}^{M}X_{r}\cos (kr)$, where $M+1$ is the
number of sites involved in the transformation (for this work, we
choose $M=100$) and $X$ stands for $S$, $D$ and $P$ correlations.
The real space correlation functions are defined as
\begin{eqnarray}
\nonumber S^{m}_{r}&\equiv & \langle \mathbf{s}^{m}_{i}
\mathbf{s}^{m}_{i+r}\rangle-\langle
\mathbf{s}^{m}_{i}\rangle\langle \mathbf{s}^{m}_{i+r}\rangle,\\
D_{r}&\equiv &\langle n_{i}n_{i+r}\rangle -\langle n_{i}\rangle
\langle n_{i+r}\rangle,\\
\nonumber P_{r}&\equiv &\langle c_{i\uparrow }c_{i\downarrow
}c_{i+r\downarrow }^{\dagger }c_{i+r\uparrow }^{\dagger
}\rangle%-\langle a_{i\uparrow }a_{i\downarrow }\rangle\langle
%a_{i+r\downarrow }^{\dagger }a_{i+r\uparrow }^{\dagger }\rangle,
\end{eqnarray}
where the spin operators associated with site $i$ is given by
$\mathbf{s}^{m}_{i}\equiv c_{i\alpha }^{\dagger }\mathbf{\sigma
}^{m}_{\alpha \beta }c_{i\beta }/2$ with $\alpha$ and $\beta
=\downarrow$, $\uparrow $ and $\mathbf{\sigma}^{m}$
($m=x$,$y$,$z$) standing for the Pauli matrices. In addition, we
also calculate $\langle c^{\dagger}_{\sigma i}c_{\sigma
i+r}\rangle$ and its fourier transform, which gives the momentum
distribution for the $\sigma$-fermion.

In Fig. 1, we present the ground state phase diagram (effective
magnetic field vs. chemical potential) for Fermi mixtures based on
the asymmetric Hubbard model at $U=-4$. In the shaded parameter
regions around the centers of the diagrams, all the lattice sites
are partially filled by both species of fermions and the effective
magnetic field $\delta\mu$ is not strong enough to induce any
population imbalance. In these regions, the ground states of the
system have either charge density wave (CDW) or singlet superfluid
(SS) as the dominant quasi-long range order, depending on the
filling factor. As the effective magnetic field increases,
population imbalance is introduced and the system could be brought
into the regions filled with solid dots as shown in the diagram.
These regions are of central interest for us, since within them
the ground state of the system is found to be of the FFLO type.
(The pairing correlation and momentum distribution characterizing
the FFLO-type states will be presented later in Fig. 3 for the
sample data points marked in the phase diagrams.) In the remaining
regions of the diagrams, at least one type of fermions has zero or
unity occupation number at all sites ($\langle n_{\sigma
i}\rangle$=0 or 1) and the system behaves as normal Fermi gas. The
green lines in Fig. 1 show the phase boundaries of the normal
states (N). (We note that our so-called normal states could be
further divided into several different categories according to the
particle occupation numbers and degree of polarization, e.g.
vacuum state, fully occupied state, fully polarized state, etc.
But as they are of limited interest, we do not distinguish them in
this work.)

By comparing the diagrams corresponding to different
$t_{\downarrow}$, we can make the following observations. First,
the parameter space supporting the FFLO states shrinks with the
increase of the imbalance in $t_{\sigma}$ or, equivalently, in the
fermion masses. This can be understood by considering the
bandwidth. A larger mass imbalance effectively leads to a narrower
bandwidth ($\sim 4t_{\downarrow}^2/U$), hence when we tune $\mu$
or $\delta\mu$ it is easier for the system to fall into the
$n_{\downarrow}=0$ or 1 bands and become normal. A second
observation is that the symmetry of the $t_{\downarrow}\ne1$
diagrams is different from that of the $t_{\downarrow}=1$ diagram.
When $t_{\downarrow}\ne1$, the diagrams are only symmetric about
the center point ($\mu=U/2$,$\delta\mu=0$), reflecting the fact
that the Hamiltonian is invariant under the particle-hole
transformation combined with the inversion about
($\mu=U/2$,$\delta\mu=0$). In the $t_{\downarrow}=1$ case, the
diagram is symmetric about the two axes $\mu=U/2$ and
$\delta\mu=0$, since the Hamiltonian now possesses an extra
symmetry, namely the invariance under spin flip combined with the
inversion about ($\mu=U/2$,$\delta\mu=0$).
\begin{figure} [tbp]
\includegraphics [height=8.5 cm, width=8 cm] {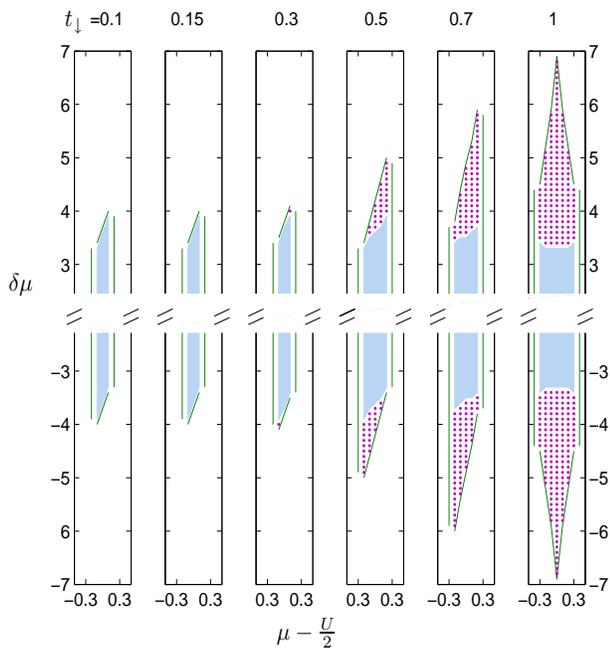}
\caption [Fig. 2] {Effective magnetic field vs. chemical potential
ground state phase diagrams for $U=-10$. From left to right, the
six panels correspond to $t_{\downarrow}=0.1$, 0.15, 0.3, 0.5,
0.7, and 1, respectively. The ground states of the system in the
regions filled with solid dots are FFLO-type states. In the shaded
regions, the fermions of different species have equal population.
In the remaining regions, the system behaves as normal Fermi gas.}
\end{figure}

We present in Fig. 2 the ground state phase diagram for $U=-10$.
One can see that the key features are the same as the $U=-4$ case.
Namely, FFLO is the only partially polarized superfluid states in
the phase diagram and its parameter space shrinks with the
increase of mass imbalance. Furthermore, the shape and symmetry of
the diagrams also resemble those of the $U=-4$ diagrams in Fig. 1.
Nonetheless, there are noticeable differences. With a stronger
on-site attraction, it is harder to break the pairs. Hence we have
higher critical fields $\delta\mu_c$ in the $U=-10$ case. Besides,
the effects of unequal masses become more dramatic when the
on-site interaction is stronger. For example, in contrast to the
$U=-4$ case, when $U=-10$ and $t_{\downarrow}=0.3$ one can barely
find the FFLO phase and at $t_{\downarrow}=0.15$ the parameter
space for FFLO completely disappears for $U=-10$.

\begin{figure} [tbp]
\includegraphics [height=8 cm, width=8 cm] {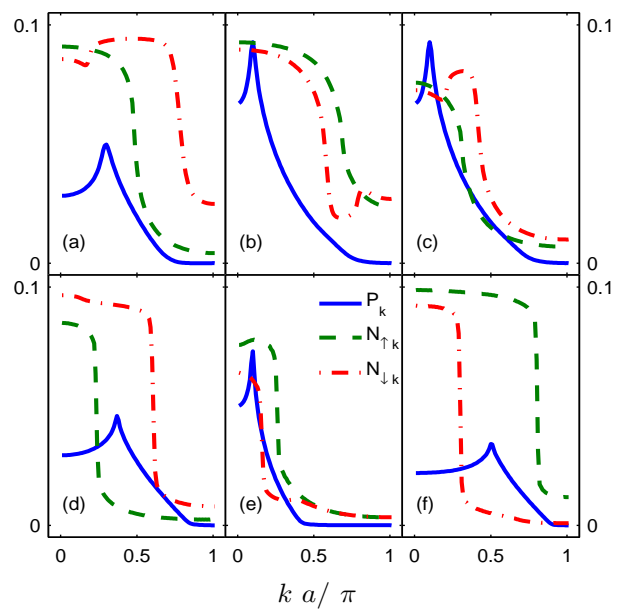}
\caption [Fig. 3] {Fourier transformations of the pairing
correlation functions ($P_k$) and particle number distributions
($N_{\uparrow k}$ and $N_{\downarrow k}$) for the partially
polarized asymmetric Hubbard model at $U=-4$. $P_k$, $N_{\uparrow
k}$, and $N_{\downarrow k}$ are depicted by the solid, dashed, and
dash-dotted curves, respectively. Other parameters
$\{t_{\downarrow},\mu,\delta\mu\}$: (a) $\{0.1,-2.2,-1.4\}$; (b)
$\{0.3,-1.7,1.4\}$; (c) $\{0.3,-2.3,-1.4\}$; (d)
$\{0.5,-2.4,-2.1\}$; (e) $\{0.5,-2.4,0.7\}$; and (f)
$\{0.7,-1.7,2.6\}$. Note that the data points corresponding to
these parameter sets are marked in Fig. 1.}
\end{figure}

In Fig. 3, we show the pairing correlation functions and particle
number distributions in the momentum space for six sample data
points with various $\{t_{\downarrow},\delta\mu,\mu\}$ in the FFLO
regime of Fig. 1. FFLO pairs are known to have non-zero
center-of-mass momentum and as its signature, the pairing
correlation function $P_k$ for FFLO states has peaks at non-zero
momentum $|k_{\uparrow F}-k_{\downarrow F}|$ with $k_{\sigma
F}\equiv\langle n_{\sigma i}\rangle\pi$ standing for the fermi
momentum for non-interacting free fermions. From the solid curves
in Fig. 3, we can clearly see the peaks of $P_k$ at non-zero
momenta $k_p=|k_{\uparrow F}-k_{\downarrow F}|$, indicating the
presence of FFLO pairing. One can also see that the particle
number distribution functions $N_{\uparrow k}$ and $N_{\downarrow
k}$ drop sharply at $k_{\uparrow F}$ and $k_{\downarrow F}$,
respectively. Another noteworthy point is that the momentum
distribution function for the ``heavier'' species
($\downarrow$-fermion) in an FFLO state clearly exhibits a dip at
momentum $2k_{\uparrow F}-k_{\downarrow F}$.

\begin{figure} [tbp]
\includegraphics [height=7 cm, width=8 cm] {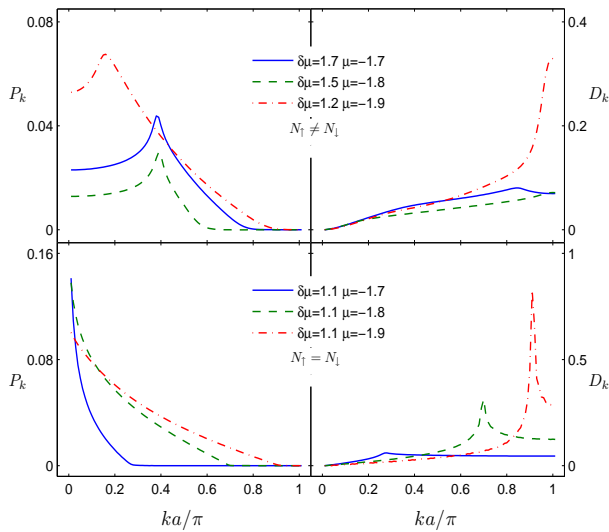}
\caption [Fig. 3] {Momentum space pairing and density correlation
functions for the asymmetric Hubbard model at $U=-4$ and
$t_{\downarrow}/t_{\uparrow}=0.15$, which is exactly the mass
ratio between $^6$Li and $^{40}$K.}
\end{figure}

Considering the potential interests in studying the mixtures of
$^6$Li and $^{40}$K experimentally\cite{Wille08}, we also present
the momentum space pairing ($P_k$) and density ($D_k$)
correlations for the asymmetric Hubbard model at $U=-4$ and
$t_{\downarrow}/t_{\uparrow}=0.15$ in Fig. 4. The two upper panels
of Fig. 4 show the pairing and density correlations in the case
with population imbalance, while the lower panels show the case
with equal populations. From the visibility and height of the
peaks in the correlation functions, one can tell which kind order
is more dominant. First, we look into the case with equal
populations. When the filling factor slightly deviates from
half-filling, charge density wave is the dominant quasi-long range
order. As an example, for $\delta\mu= 1.1$ and $\mu=-1.9$, we have
$\langle n_i \rangle=2k_{\uparrow F}=2k_{\downarrow F}\sim 1.09$.
One can observe a sharp peak for $D_k$ locating at
$k\sim0.9\pi(=2\pi-2k_{\sigma F})$ while $P_k$ shows a much
broader and lower peak at $k=0$ indicating that CDW is more
dominant than SS. When the filling factor is further away from
half-filling, the singlet superfluid order gradually becomes more
dominant. (The peak of $D_k$ moves toward $k=0$ with its
visibility and height decreasing, while the peak of $P_k$ remains
at $k=0$ and becomes sharper.) Next, we examine the case with
population imbalance. From the upper left panel of Fig. 4, we can
see that $P_k$ shows maxima at non-zero momenta, which are
verified to be directly given by the difference in particle number
density, signalling the presence of FFLO pairing. The results
presented in Fig. 4 indicate that in order to observe the FFLO
state in mixtures of spinless $^6$Li and $^{40}$K, one should keep
the density of $^{40}$K well away from half-filling and consider
avoid large inter-species interaction.

In summary, we have presented effective magnetic field
($\delta\mu$) vs. chemical potential ($\mu$) phase diagrams for 1D
fermi mixtures with unequal masses and attractive on-site
inter-species interactions loaded in optical lattices. We find
that with mass and population imbalance the ground state of the
system could be either an FFLO state or a normal state. With the
increase of mass imbalance, the parameter space for the FFLO
states shrinks. When the mass imbalance gets too large, the FFLO
states are no longer stable.

We acknowledge helpful comments from A. Feiguin and W. V. Liu.
This work is supported by ARO-DARPA.


\begin{thebibliography}{99}
\bibitem{FF64} P. Fulde, and R. A. Ferrell, \textit{Phys. Rev.} \textbf{135},
A550 (1964).

\bibitem{LO65} A. I. Larkin, and Yu. N. Ovchinnikov, \textit{Sov. Phys. JETP} \textbf{20}, 762 (1965).

\bibitem{Sarma63} G. Sarma, \textit{J. Phys. Chem.
Solids}, \textbf{24}, 1029 (1963).

\bibitem{Liu03} W.V. Liu and F. Wilczek, \textit{Phys. Rev. Lett.} \textbf{90}, 047002
(2003). %BP2, unequal mass

\bibitem{Zwierlein06} M. W. Zwierlein \textit{et al.},
\textit{Science} \textbf{311}, 492 (2006); M. W. Zwierlein
\textit{et al.}, \textit{Nature} \textbf{442}, 54 (2006); Y. Shin
\textit{et al.}, \textit{Phys. Rev. Lett.} \textbf{97}, 030401
(2006); Y. Shin \textit{et al.}, \textit{Nature} 451, 689 (2008);
G. B. Partridge \textit{et al.}, \textit{Science} \textbf{311},
503 (2006); G. B. Partridge \textit{et al.}, \textit{Phys. Rev.
Lett.} \textbf{97}, 190407
(2006). %exp
%%%%%%%%%%%%%%%%%%%%5

\bibitem{3DFFLO} T. Mizushima \textit{et al.}, \textit{Phys. Rev. Lett.} \textbf{94},
060404 (2005); D. E. Sheehy and L. Radzihovsky, \textit{ibid.}
\textbf{96}, 060401 (2006); J. Kinnunen \textit{et al.},
\textit{ibid.} \textbf{96}, 110403 (2006); K. Machida \textit{et
al.}, \textit{ibid.} \textbf{97}, 120407 (2006); P. Castorina
\textit{et al.}, \textit{Phys. Rev. A} \textbf{72}, 025601 (2005);
N. Yoshida and S.-K. Yip, \textit{ibid.} \textbf{75}, 063601
(2007); W. Zhang and L.-M. Duan, \textit{Phys. Rev. A}
\textbf{76}, 042710 (2007); T. K. Koponen, \textit{et al.},  \textit{New J. Phys.} \textbf{10}, 045014 (2008); %3D FFLO population imbalance



\bibitem{3DSarma} K. B. Gubbels \textit{et al.}, \textit{Phys. Rev. Lett.} \textbf{97}, 210402
(2006); T.-L. Dao \textit{et al.}, \textit{ibid.} \textbf{101},
236405 (2008); C.-C. Chien \textit{et al.}, \textit{ibid.}
\textbf{97}, 090402 (2006); \textit{ibid.} \textbf{98}, 110404
(2007).
%population imbalance Sarma;


%\bibitem{Pilati08} S. Pilati and S. Giorgini, \textit{Phys. Rev. Lett.} \textbf{100}, 030401 (2008).%3D phase separation, no claim for FFLO or BP

\bibitem{1DFFLO} K. Yang, \textit{Phys. Rev. B} \textbf{63}, 140511(R) (2001); E. Zhao and W. V. Liu, \textit{Phys. Rev. A} \textbf{78}, 063605
(2008); M. M. Parish \textit{et al.}, \textit{Phys. Rev. Lett.}
\textbf{99}, 250403 (2007); H. Hu \textit{et al.}, \textit{ibid.}
\textbf{98}, 070403 (2007); G. Orso, \textit{ibid} \textbf{98},
070402
(2007). % FFLO quasi-1D, 1D : mean field, bosonization, bethe ansatz

\bibitem{1DFFLOb} A. Feiguin and F. Heidrich-Meisner, \textit{Phys. Rev. B} \textbf{76}, 220508(R)
(2007); M. Tezuka and M. Ueda, \textit{Phys. Rev. Lett.}
\textbf{100}, 110403 (2008); M. Rizzi \textit{et al.},
\textit{Phys. Rev. B} \textbf{77}, 245105 (2008); G. G. Batrouni
\textit{et al.}, \textit{Phys. Rev. Lett.} \textbf{100}, 116405
(2008);A. L\"{u}scher \textit{et al.}, \textit{Phys. Rev. A}
\textbf{78}, 013637 (2008); A. Feiguin and David A. Huse, arXiv:0809.3024; A. Feiguin and F. Heidrich-Meisner, arXiv:0809.1539. %DMRG, QMC

%%%%%%%%%%%%%%%%%%%%%%%%%%%%%%%%%%%%%%%%%%%%%%%%%%%%%%%%%%%%%%%%%%%%%%%%%%%%%%

\bibitem{Cazalilla05} M. A. Cazalilla \textit{et al.}, \textit{Phys. Rev. Lett.} \textbf{95}, 226402 (2005). %renormalization group equal spin population, phase diagram, state-dependent lattice

\bibitem{Gu07} S.-J. Gu \textit{et al.}, \textit{Phys. Rev. B} \textbf{76}, 125107 (2007).%DMRG, AHM, unpolarized

\bibitem{3Dmass} D. S. Petrov \textit{et al.}, \textit{J. Phys. B} \textbf{38},
S645 (2005); M. Iskin and C. A. R. S\'{a} de Melo, \textit{Phys.
Rev. Lett.} \textbf{97}, 100404 (2006); G. Orso \textit{et al.},
arXiv:0709.1690; D. S. Petrov \textit{et al.}, \textit{Phys. Rev.
Lett.} \textbf{99}, 130407 (2007); M. M. Parish \textit{et al.},
\textit{Phys. Rev. Lett.} \textbf{98}, 160402 (2007); S. A.
Silotri \textit{et al.}, arXiv:0805.1784; G.-D. Lin \textit{et
al.}, \textit{Phys. Rev. A} \textbf{74},
031604(R) (2006).%unequal mass

\bibitem{Batrouni08b} G. G. Batrouni \textit{et al.}, arXiv:0809.4549. % QMC 1D FFLO, unequal mass

\bibitem{Liu04} W. V. Liu \textit{et al.}, \textit{Phys. Rev. A} \textbf{70}, 033603 (2004).%spin-dependent Hubbard model with cold atoms in optical lattice

\bibitem{TEBD} G. Vidal, \textit{Phys. Rev. Lett.} \textbf{91}, 147902
(2003); \textit{ibid.} \textbf{93}, 040502 (2004); \textit{ibid.}
\textbf{98}, 070201 (2007).

\bibitem{Wille08} E. Wille \textit{et al.}, \textit{Phys. Rev. Lett.} \textbf{100}, 053201 (2008); A.-C. Voigt \textit{et al.}, \textit{ibid.} \textbf{102}, 020405 (2009).

\end{thebibliography}
\end{document}